\documentclass[preprint,aps]{revtex4-1}

\usepackage{amsmath,amssymb,amsfonts,dcolumn,color,graphicx,graphics,latexsym,placeins,epsfig}
\usepackage{subfigure}
\usepackage{natbib}
\newcommand{\be}{\begin{equation}}
\newcommand{\ee}{\end{equation}}
\newcommand{\ba}{\begin{eqnarray}}
\newcommand{\ea}{\end{eqnarray}}

\begin{document}

\title{\Large \bf Born-Infeld cosmology with scalar Born-Infeld matter}

\author{Soumya Jana and Sayan Kar}
\email{soumyajana@phy.iitkgp.ernet.in, sayan@iitkgp.ac.in}
\affiliation{\rm Department of Physics {\it and} Centre for Theoretical Studies \\Indian Institute of Technology, Kharagpur, 721302, India}

\begin{abstract}
\noindent Cosmology in Eddington-inspired Born-Infeld gravity is
investigated using a scalar Born-Infeld field (e.g. tachyon condensate) 
as matter. In this way, both in the gravity and matter sectors 
we have Born-Infeld-like structures characterized by their
actions and via two separate constants, $\kappa$ and $\alpha_T^2$ respectively. 
With a particular choice of the form of $\dot{\phi}$ (the time derivative of the 
Born-Infeld scalar), analytical cosmological solutions are found.
Thereafter, we explore some of the unique features of the
corresponding cosmological spacetimes. For
$\kappa>0$, our solution has a de Sitter-like expansion both at early and 
late times, with an intermediate deceleration sandwiched between the
accelerating phases. On the other hand, when $\kappa<0$, 
the initial de Sitter phase is replaced by a  bounce. 
Our solutions, at late time, fit well with available supernova data-- a fact 
we demonstrate explicitly. The estimated properties of the Universe obtained 
from the fitting of the $\kappa>0$ solution, are as good 
as in $\Lambda$CDM cosmology. However, the $\kappa<0$ solution has to be 
discarded due to the occurrence of a bounce at an unacceptably 
low redshift.  
\end{abstract}


\maketitle

\section{\bf Introduction} 

\noindent General relativity (GR) is largely 
successful as a classical theory of gravity, especially with the
recent detection of gravitational waves. However,  
there do exist many unsolved puzzles.
Among them is the problem of singularities in GR which is expected to
be resolved by quantum gravity. Dark matter and dark energy also do not seem
to be well understood within the framework of GR. 
Thus, in order to address some of the above-mentioned problems, 
it is not unusual to construct theories which deviate from GR within the
classical framework, inside matter distributions, 
or in the strong-field regime. This expectation has 
led many researchers to actively pursue 
modified theories of gravity in the classical domain 
and also in quantum theory. 
One such modification is inspired by the well-known 
Born-Infeld electrodynamics where we are able to regularize the infinity 
in the electric field at the location of a point charge \cite{born}. 
With a similar determinantal structure $\left[\sqrt{-det(g_{\mu\nu}+\kappa 
R_{\mu\nu})} \right]$ as in the action of Born-Infeld electrodynamics, 
a gravity theory in the metric formulation was first suggested by 
Deser and Gibbons \cite{desgib}. 
In fact, the determinantal form of the gravitational action existed earlier in Eddington's re-formulation of GR in de Sitter spacetime 
\cite{edd}. This formulation is affine and the connection is the basic variable 
instead of the metric. However, the coupling of matter remained a problem 
in Eddington's approach.

\noindent Later, Vollick \cite{vollick} introduced the Palatini formulation of  
Born-Infeld gravity and worked on various related aspects. 
He also introduced a nontrivial and somewhat artificial way of coupling 
matter in such a theory \cite{vollick2,vollick3}. More recently, 
Banados and Ferreira  \cite{banados} have come up with a formulation 
where the matter coupling is 
different and simpler compared to Vollick's proposal. 
We will focus here on the theory proposed in Ref. \cite{banados} and refer to it
as Eddington-inspired Born--Infeld (EiBI) gravity, for obvious reasons. 
Note that the EiBI theory has the feature that it reduces to GR in vacuum.

\noindent Interestingly, EiBI theory also falls within the class of bimetric 
theories of gravity (also called bi-gravity). The  current bimetric theories 
have their origin in the seminal work of Isham, Salam and Strathdee \cite{isham}. Several articles have appeared in the last few years on various aspects of 
such bi-gravity theories. In Ref. \cite{scargill}, the authors pointed out that the EiBI field equations can also be derived from an equivalent bi-gravity 
action. This action is closely related to a recently discovered family of 
unitary massive gravity theories which are built as bi-gravity theories. 
Several others  have contributed in this direction, in various ways 
\cite{schmidt,lavinia}.   

\noindent Let us now briefly recall Eddington--inspired Born--Infeld gravity. 
The central feature here is the existence of a physical metric which couples to matter and another auxiliary metric which is not used for matter couplings. 
One needs to solve for both metrics through the field equations. The action for the theory developed in Ref. \cite{banados} is given as
\begin{equation}
S_{BI}(g,\Gamma, \Psi) =\frac{c^3}{8\pi G\kappa}\int d^4 x \left [ \sqrt{-\vert g_{\mu\nu} +\kappa R_{\mu\nu}(\Gamma)\vert}-\lambda \sqrt{-g} \right]+ S_M (g, \Psi)
\end{equation}
where $\Lambda= \frac{\lambda-1}{\kappa}$. 
 Variation with respect to $\Gamma$, done using the
auxiliary metric $q_{\mu\nu}=g_{\mu\nu}+\kappa R_{\mu\nu}(q)$, gives
\begin{equation}
q_{\mu \nu}=g_{\mu\nu} + \kappa R_{\mu\nu}(q)
\label{eq:gammavarn}
\end{equation}
Variation with respect to $g_{\mu\nu}$ gives 
\begin{equation}
\sqrt{-q} q^{\mu\nu} = \lambda \sqrt{-g}g^{\mu\nu}-\frac{8\pi G}{c^4}\kappa \sqrt{-g} T^{\mu\nu}
\label{eq:gvarn}
\end{equation}
where the $T^{\mu\nu}$ components are in the coordinate frame.
In order to obtain solutions, we need to assume a $g_{\mu\nu}$ and 
a $q_{\mu\nu}$
with unknown functions, as well as a matter stress-energy ($T^{\mu\nu}$). 
Thereafter, we write down the field equations and obtain solutions using 
some additional assumptions about the metric functions and the stress-energy. 

\noindent Work on various fronts has been carried out
on various aspects of this theory in the last few years.
Astrophysical scenarios have been discussed  in Refs. \cite{cardoso,casanellas,avelino,sham,sham2,structure.exotic.star,sotani.neutron.star,
sotani.stellar.oscillations,sotani.magnetic.star}. Spherically symmetric solutions were obtained in Refs. 
\cite{banados,wei,sotani,eibiwormhole,rajibul,jana2}.
A domain wall brane in a higher-dimensional generalization
was analyzed in Ref. \cite{eibibrane}. 
Generic features of the paradigm of matter-gravity couplings were analyzed in Ref. \cite{delsate}. In Ref. \cite{cho_prd88}, authors showed that EiBI theory admits a nongravitating matter distribution, which is not allowed in GR.  Some interesting cosmological and circularly symmetric solutions in $2+1$ dimensions were shown in Ref. \cite{jana}. 
Constraints on the EiBI parameter $\kappa$ have been obtained  
from studies on compact stars \cite{cardoso,sham2}, tests in the 
Solar System \cite{solar.test}, astrophysical and cosmological observations \cite{avelino}, and nuclear physics studies \cite{nuclear.test}. 
In Ref. \cite{pani}, a major problem related to surface singularities
was noticed in the context of  stellar physics. However, gravitational 
backreaction was suggested as a cure to this problem 
in Ref. \cite{eibiprob.cure}. In Ref. \cite{odintsov}, the authors proposed a modification to EiBI theory by taking its functional extension in a way similar to $f(R)$ theory. Recently, the authors of Ref. \cite{fernandes} used a different way of matter 
coupling by taking the Kaluza ansatz for the five-dimensional EiBI action 
in a purely metric formulation, and then compactify it using Kaluza's procedure to get a four-dimensional gravity coupled in a nonlinear way to 
electromagnetic theory. 


\noindent Much work in EiBI gravity is devoted to cosmology. 
In Refs. \cite{banados,scargill,cho}, the authors showed the nonsingularity 
of the Universe filled by any ordinary matter. Linear perturbations have 
been studied in the background of the homogeneous and isotropic spacetimes 
in the Eddington regime \cite{escamilla,linear.perturbation}. Bouncing 
cosmology in EiBI gravity was emphasized as an alternative to  inflation 
in Ref. \cite{avelinoferreira}. The authors of Ref. \cite{chokim} studied a model 
described by a scalar field with a quadratic potential, which results in a 
nonsingular initial state of the Universe leading naturally to inflation. 
They also investigated the stability of the tensor perturbations in this 
inflationary model \cite{cho90} while the scalar perturbation was studied 
in Ref. \cite{cho_scalar_perturbation}. Other relevant work has been reported in 
Refs. \cite{cho_spectral_indices,cho_pow_spectra,cho_tensor_scalar}. 
Large-scale structure formation in the Universe and the integrated 
Sachs-Wolfe effect are discussed in Ref. \cite{large.scale.structure}. 
Further efforts in this line were reported in Refs. \cite{felice,power.spectrum,cascading_dust_inflation,bianchi.cosmo}. 

\noindent In Ref. \cite{jana2}, we considered the Born-Infeld structure in both 
the gravity and matter sectors and obtained new spherically symmetric static 
spacetimes when EiBI gravity is coupled to Born-Infeld electrodynamics. 
In this article, we investigate a similar problem in EiBI cosmology where we 
consider a Born-Infeld scalar field in the matter part. In particular, we use 
the tachyon condensate scalar because of  its Born-Infeld-like structure 
in the action. The tachyon condensate scalar field arises in the context of 
theories of unification such as superstring theory \cite{tachyon_asen}. There 
have been several articles in the literature where cosmology with the tachyon 
scalar field was discussed \cite{tachyon_cosmo1,tachyon_cosmo2,tachyon_cosmo3,tachyon_cosmo4,tachyon_cosmo5,tachyon_cosmo6,
tachyon_cosmo7,tachyon_cosmo8,tachyon_cosmo9,tachyon_cosmo10,tachyon_cosmo11,tachyon_cosmo12,
tachyon_cosmo13}. 
The energy-momentum tensor of the BI scalar tachyon condensate can be 
split into two parts; one with zero pressure (dark matter) and another 
with $p=-\rho$ (dark energy) \cite{padmanabhan_tachyon1}. This facilitates 
the description of dark energy and dark matter using a single scalar field, 
a fact we shall use for our model in this article. 

\noindent We organize our article as follows. 
In Sec. II we discuss the basic setup of the BI scalar field in the 
cosmological background. Next, in Sec. III we obtain a constant 
negative (effective) pressure solution as the special case of the general 
assumption on the form of the time derivative of the scalar field. Then, 
we fit the solutions with the supernova data and test its viability 
in Sec. IV. Finally, in Sec. V we summarize our results.                        

\section{Cosmology with a Born-Infeld scalar}
\label{sec:cosmo}
The action for the BI scalar (eg. tachyon condensate)is
\begin{equation}
S_M=-\frac{1}{c}\int \sqrt{-g}\, \alpha_T^2\mathcal{V}(\phi)\sqrt{1+\alpha^{-2}_Tg^{\mu\nu}\partial_{\mu}\phi\partial_{\nu}\phi}\, d^4x
\label{eq:tachy_scalar_lag}
\end{equation} 
where, $\mathcal{V}(\phi)$ is the potential for the scalar field and $\alpha_T$ is the constant parameter. The resulting stress-energy tensor components 
have the following general expression:
\begin{eqnarray}
T^{\mu\nu}&=&\frac{2}{\sqrt{-g}}\frac{\partial \mathcal{L}}{\partial g_{\mu\nu}}\nonumber\\
&=& \mathcal{V}(\phi)\left[\frac{g^{\mu \alpha}g^{\nu \beta}\partial_{\alpha}\phi \partial_{\beta}\phi -g^{\mu\nu}g^{\alpha\beta}\partial_{\alpha}\phi\partial_{\beta}\phi -g^{\mu\nu}\alpha_T^2}{\sqrt{1+\alpha^{-2}_Tg^{\alpha\beta}\partial_{\alpha}\phi\partial_{\beta}\phi}}\right]
\label{eq:scalar_stress_energy}
\end{eqnarray} 
By varying the action with respect to (w.r.t.) $\phi$, we get the equation of motion of the scalar field
\begin{equation}
\partial_{\nu}\left[\frac{\mathcal{V}(\phi)\sqrt{-g}g^{\mu\nu}\partial_{\mu}\phi}{\sqrt{1+\alpha^{-2}_Tg^{\alpha\beta}\partial_{\alpha}\phi\partial_{\beta}\phi}} \right]=\alpha_T^2\sqrt{-g}\mathcal{V}'(\phi)\sqrt{1+\alpha^{-2}_Tg^{\alpha\beta}\partial_{\alpha}\phi\partial_{\beta}\phi}
\label{eq:eom_scalar}
\end{equation} 
where $\mathcal{V}'(\phi)$ is derivative of the potential w.r.t. 
the scalar field. The equation~(\ref{eq:eom_scalar}) also ensures the 
conservation of the stress-energy tensor ($\nabla_{\mu}T^{\mu\nu}=0$). 
For a homogeneous and isotropic, spatially flat Universe, we assume the 
following ansatz 
for the physical spacetime metric:
\begin{equation}
ds^2=-U(t)c^2dt^2+a^2(t)\left[dx^2+dy^2+dz^2\right]
\label{eq:phy_metric_ansatz}
\end{equation}  
The equation~(\ref{eq:eom_scalar}) leads to the following equation of motion of the scalar field ($\phi$):
\begin{equation}
\frac{\ddot{\phi}}{c^2\alpha^2_TU-\dot{\phi}^2}+\frac{3\dot{\phi}}{c^2\alpha^2_TU}\left(\frac{\dot{a}}{a}\right)+\frac{\mathcal{V}'(\phi)}{\mathcal{V}(\phi)}-\frac{\dot{\phi}\dot{U}}{2U(c^2\alpha_T^2U-\dot{\phi}^2)}=0
\label{eq:eom_scalar_cosmo}
\end{equation} 
where dots denote the derivatives w.r.t. $t$ and primes denote the 
derivatives w.r.t. $\phi$. Also, the stress-energy tensor for the BI 
scalar field can be re-written as that of an analogous perfect fluid, 
$i.e.$ $T_{\mu\nu}=\left(p_{\phi}+\rho_{\phi}c^2\right)u_{\mu}u_{\nu}+p_{\phi}g_{\mu\nu}$, where $p_{\phi}$ and $\rho_{\phi}$ are the equivalent pressure and energy density, respectively, in the comoving frame. 
The $\rho_\phi$ and $p_\phi$ are expressed in terms of the scalar field as
\begin{eqnarray}
\rho_{\phi}&=&\frac{\alpha_T^2\mathcal{V}(\phi)}{c^2\sqrt{1-\dot{\phi}^2U^{-1}\alpha_T^{-2}c^{-2}}}\label{eq:rho_phi}\\
p_{\phi}&=&-\alpha_T^2\mathcal{V}(\phi)\sqrt{1-\dot{\phi}^2U^{-1}\alpha_T^{-2}c^{-2}}
\label{eq:p_phi}
\end{eqnarray} 
We can re-express the scalar potential in terms of 
$\rho_{\phi}$ and $p_{\phi}$ in the following way,
\begin{equation}
\mathcal{V}(\phi)=\sqrt{-p_{\phi}\rho_{\phi}c^2}/\alpha_T^2
\label{eq:pot_tach}
\end{equation} 
Using Eqs.~(\ref{eq:rho_phi}) and (\ref{eq:p_phi}), Eq.~(\ref{eq:eom_scalar_cosmo}) can be rewritten as
\begin{equation}
\frac{\dot{\rho}_{\phi}}{\rho_{\phi}}=-3\left(\frac{\dot{a}}{a}\right)\left(\frac{\dot{\phi}^2}{Uc^2\alpha_T^2}\right)
\label{eq:rhodot}
\end{equation} 
Generally, in a scalar field cosmology, we choose the form of the potential 
for the scalar field. This is because we do have an extra degree of freedom 
in the field equations. Here, we exploit this and choose the form of $\dot{\phi}$ instead of $\mathcal{V}(\phi)$. We assume the form of $\dot{\phi}$ as
\begin{equation}
\dot{\phi}^2=\frac{Uc^2\alpha_T^2}{1+C_1a^n}
\label{eq:phidot_assume}
\end{equation}
where $C_1>0$ and $n>0$. Thus, using the Eq.~(\ref{eq:phidot_assume}) we 
solve Eq.~(\ref{eq:rhodot}) and obtain
\begin{eqnarray}
\rho_{\phi}&=&C_2(a^{-n}+C_1)^{3/n}\label{eq:rho_a}\\
p_{\phi}&=&-C_1C_2c^2(a^{-n}+C_1)^{3/n-1}
\label{eq:p_a}
\end{eqnarray}
where $C_2$ is an integration constant. At this point, $\alpha_T$ is just a
scaling parameter for the scalar field and the potential. But this is changed 
if we relate $C_1$ with $\alpha_T$. We choose $C_1^{3/n}=\alpha_T^{2}$. Then the potential becomes [from Eq.~(\ref{eq:pot_tach})]
\begin{equation}
\mathcal{V}=C_2c^2\left(a^{-n}/\alpha_T^{2n/3}+1\right)^{3/n-1/2}
\label{eq:pot_a}
\end{equation}
where the potential is now expressed in terms of the scale factor 
in parametrized form. For this choice of $C_1$, $\mathcal{V}\rightarrow C_2 c^2$ for $a\rightarrow \infty$. This becomes a universal character ($i.e.$ for arbitrary $n>0$). For $a\rightarrow \infty$, $\rho_{\phi}\rightarrow \alpha_T^2C_2$ and $p_{\phi}\rightarrow -\alpha_T^2C_2c^2$. On the other hand, 
for the given $n$ and $C_2$, $\alpha_T$ becomes the control parameter 
for modifying the behaviour of the potential and all other quantities 
in the early Universe.  
   
\section{ A constant negative pressure solution}
In GR, the cosmological solution with a constant negative pressure can be found. It may be a bouncing solution where the scale factor is given by
\begin{equation}
a(t) =a_0 \left ( \cosh \frac{t}{t_0}\right )^{\frac{2}{3}}
\end{equation}
The corresponding energy density profile is
\begin{equation}
\rho=\frac{1}{6\pi G t_0^2}\left(1-\frac{a_0^3}{a^3} \right)
\end{equation}
The other possibility is of a singular solution where the scale factor and the energy density are given by
\begin{eqnarray}
a(t) =a_0 \left ( \sinh \frac{t}{t_0}\right )^{\frac{2}{3}}\\
\rho=\frac{1}{6\pi G t_0^2}\left(1+\frac{a_0^3}{a^3} \right)
\end{eqnarray}
 
\noindent In EiBI gravity coupled to a BI scalar, with the choice of $n=3$ 
in Eq.~(\ref{eq:p_a}), $p_{\phi}= -\alpha_T^2C_2 c^2$ (constant negative pressure). We would now like to construct the solution for the physical line element.
Let us assume an ansatz for the auxiliary metric 
\begin{equation}
ds^2_q= -V(t)c^2dt^2+b^2(t)\left[dx^2+dy^2+dz^2\right]
\label{eq:auxiliary_metric_ansatz}
\end{equation}
From the Eq.~(\ref{eq:gvarn}) we get two independent equations which we 
rewrite in the following way,
\begin{eqnarray}
a&=& \frac{b}{C_0}\sqrt{V/U} ,\label{eq:a_bUV}\\
\rho_{\phi}&=& \frac{c^2}{8\pi G \kappa}\left(\frac{C_0^3U^2}{V^2}-1\right)
\label{eq:rhophi_a}
\end{eqnarray}
where $C_0=1+8\pi G \kappa \alpha_T^2 C_2 c^{-2}$ is a new constant parameter.
Note that $C_0$ has both $\kappa$ and $\alpha_T^2$ in its definition.
We will see that $C_0$ is one of the parameters which we will use while 
fitting our final solution with supernova observations.
We also assume $\lambda=1$ ($\Lambda=0$).
From the $\Gamma$ variation, we further  have  two independent equations 
[Eq.~(\ref{eq:gammavarn})]. After a simple algebraic manipulation 
these two equations become
\begin{eqnarray}
\frac{\dot{b}^2}{b^2}=\frac{c^2}{6\kappa}\left( 2V+U-\frac{3V^2}{C_0^2U}\right) \label{eq:bdotequation}\\
\frac{d}{dt}\left(\frac{\dot{b}}{b}\right)-\frac{1}{2}\frac{\dot{b}\dot{V}}{bV}=\frac{c^2}{2\kappa}\left(-U+\frac{V^2}{C_0^2U}\right)
\label{eq:bddot}
\end{eqnarray}
To obtain the solution, we need to solve the four equations~(\ref{eq:a_bUV}), (\ref{eq:rhophi_a}), (\ref{eq:bdotequation}), and (\ref{eq:bddot}). 
However, we have five unknowns: $a, b, U, V,$ and $\rho_{\phi}$. Therefore, 
we have the freedom to choose the functional form of any one of the unknowns. 
Since we are free to choose the auxiliary metric functions 
we assume $b(t)=b_0\exp(\bar{H}_b t)$, where $b_0$ and $\bar{H}_b$ are two 
arbitrary nonzero constants. Thus, from Eq.~(\ref{eq:bdotequation}), we 
arrive at
\begin{equation}
U^2+2\left(V-\frac{3\kappa \bar{H}_b^2}{c^2}\right)U-\frac{3V^2}{C^2_0}=0
\label{eq:U_square}
\end{equation} 
The above quadratic equation~(\ref{eq:U_square}) has two roots. We choose 
the one for which $U>0$. Therefore, we have 
\begin{equation}
U=\sqrt{\left(V-\frac{3\kappa \bar{H}_b^2}{c^2}\right)^2+\frac{3V^2}{C_0^2}}-\left(V-\frac{3\kappa \bar{H}_b^2}{c^2}\right)
\label{eq:U_root}
\end{equation} 
Further, from Eq.~(\ref{eq:bddot}), we have
\begin{equation}
\frac{\dot{V}}{V}=\frac{c^2}{\bar{H}_b\kappa}\left(U-\frac{V^2}{C_0^2U}\right)
\label{eq:Vdot}
\end{equation}
At this juncture, we introduce a new variable $X= C_0\left(1-3\kappa \bar{H}_b^2c^{-2}V^{-1}\right)$ and rewrite the Eq.~(\ref{eq:Vdot}) as
\begin{equation}
\frac{\dot{X}}{2X-\sqrt{X^2+3}}= -2\bar{H}_b
\label{eq:X_dot}
\end{equation}
Integrating the Eq.~(\ref{eq:X_dot}), we get
\begin{equation}
\frac{(2X-\sqrt{X^2+3})^2}{\sqrt{X^2+3}-X}=\frac{C_3}{b^6}
\label{eq:X_b}
\end{equation}
where $C_3$ is the constant of integration. Using Eqs.~(\ref{eq:a_bUV}), (\ref{eq:U_square}), and (\ref{eq:X_b}), we compare the two expressions of $\rho_{\phi}$--one from field equations ($i.e.$ Eq.~\ref{eq:rhophi_a}) and the other 
from the conservation equation ($i.e.$ Eq.~\ref{eq:rho_a}). Since the 
field equations in EiBI theory satisfy the conservation equation, the constant 
$C_3$ is fixed and has the expression: $C_3=144\pi^2G^2\kappa^2C_2^2C_0c^{-4}$ .  
\noindent The new variable $X$ is related to $V$. Since $V$ appears in the
auxiliary metric as a coefficient of $dt^2$, $X$ is, in a sense related to
a redefinition of the time variable. In what follows, we will see this
connection more explicitly.

\subsection{$\kappa>0$} 
\begin{figure}
\centering
\includegraphics[width=0.7\textwidth]{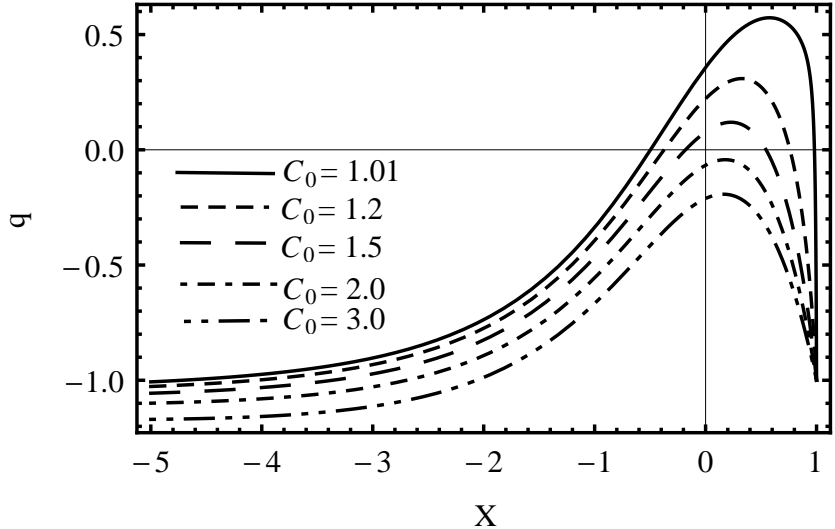}
\caption{The deceleration parameter ($q$) is plotted as function of $X$. In the plot, $X$ ranges from $-\infty$ to 1 for $a\in (0 ,\infty)$. $C_0$ [see Eq.~(\ref{eq:q_X})] takes different values for different curves in the plot. 
For $C_0 =1.01, 1.2, 1.5$, $q$ has a transition from negative to positive
values and then again to a negative value. }
\label{fig:q_c0}
\end{figure}

Using Eqs.~(\ref{eq:a_bUV}), (\ref{eq:U_root}), and (\ref{eq:X_b}) we 
rewrite the expression of $a$ as a function of $X$:
\begin{equation}
a^3= \frac{4\pi G \kappa C_2}{C_0 c^2}\left[ \frac{\sqrt{X^2+3}+X}{\sqrt{X^2+3}-2X}\right]
\label{eq:a_X}
\end{equation}
From Eq.~(\ref{eq:rho_a}), we note that $\rho_{\phi}=\alpha_T^2C_2(a^{-3}\alpha_T^{-2}+1)$ for $n=3$ and $C_1=\alpha_T^2$. So, $\rho_{\phi}>0$ necessitates $a^3>0$. Then, for $\kappa >0 $, $X \leq 1$ and $a\in (0,\infty)$ maps onto $X\in (-\infty,1)$. Similarly, we express the other metric function $U$ and the 
coordinate $t$ as functions of $X$
\begin{eqnarray}
U&=&\frac{3\kappa \bar{H}_b^2}{c^2}\left[\frac{\sqrt{X^2+3}-X}{C_0-X} \right] \label{eq:U_x}\\
t&=& t_0 +\frac{1}{6\bar{H}_b}ln\left[\frac{C_3\left(\sqrt{X^2+3}-X \right)}{\left(2X-\sqrt{X^2+3}\right)^2} \right] \label{eq:t_X}
\end{eqnarray}
where $t_0=-ln( b_0)/\bar{H}_b$ is an arbitrary constant. 
The relation between $t$ and $X$ in Eq.~(\ref{eq:t_X}) is invertible and we 
may write $X$ as a function of $t$ though the expression takes a 
complicated form. Therefore, it is better to express the metric functions 
$a(t)$ and $U(t)$ of the physical metric [Eq.~(\ref{eq:phy_metric_ansatz})] in 
parametric form [Eqs.~(\ref{eq:a_X}), (\ref{eq:U_x}), and (\ref{eq:t_X})], 
where $X$ is the parameter. We define the cosmological time $\tau$ 
\begin{eqnarray}
\tau &=& \int \sqrt{U}dt \nonumber \\
&=& \frac{\sqrt{3\kappa}}{2c}\int \left[\frac{\sqrt{X^2+3}-X}{C_0-X} \right]^{1/2}\frac{dX}{\left(\sqrt{X^2+3}-2X\right)} + const.
\label{eq:tau_X}
\end{eqnarray}
In order to understand the evolution of $a$, 
we compute the deceleration parameter ($q$)
\begin{eqnarray}
q&=& -a\frac{d^2a}{d\tau^2} \Bigg/ \left(\frac{da}{d\tau}\right)^2 \nonumber \\
&=&-1+\left(\frac{\sqrt{X^2+3}-2X}{\sqrt{X^2+3}-X}\right)\left[1+\frac{X}{\sqrt{X^2+3}}+\frac{\sqrt{X^2+3}+X-C_0}{2(C_0-X)}\right]
\label{eq:q_X}
\end{eqnarray}
Interestingly, we note that, for $X=1$ ($i.e.$ $a\rightarrow \infty$), $q= -1$. Also, for $X\rightarrow -\infty$ ($i.e.$ $a\rightarrow 0$), $q\rightarrow -1$. 
Therefore, both at early and late times, the Universe undergoes a 
de Sitter phase in our model. Also, for some values of $C_0$, the 
Universe may undergo a decelerated expansion state ($q>0$) [see Fig.~\ref{fig:q_c0}].   

\begin{figure}[!htbp]
\centering
\subfigure[]{\includegraphics[scale=0.9]{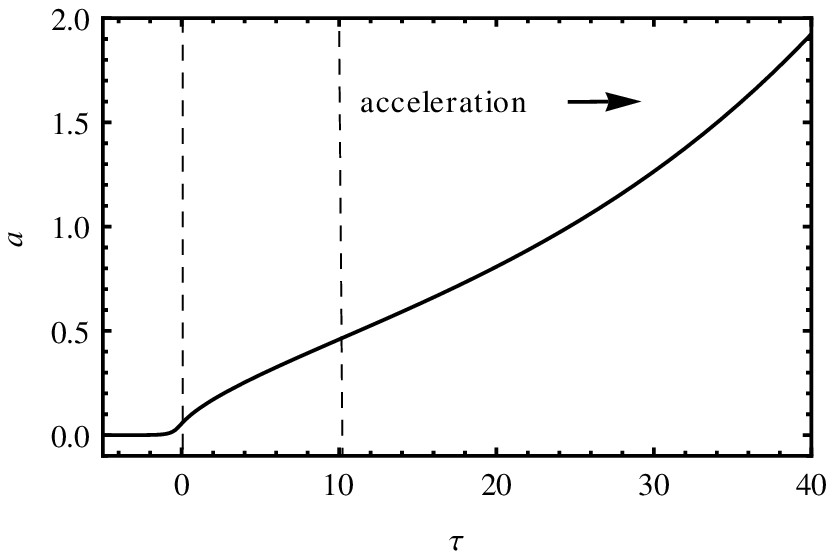}\label{subfig:a_kappapos}}
\subfigure[]{\includegraphics[scale=0.9]{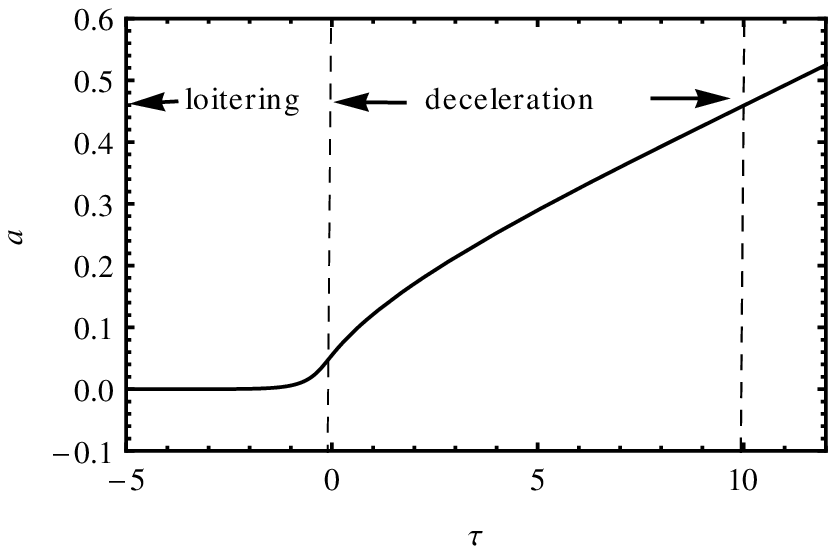}\label{subfig:a_kappapos_zoom}}\\
\subfigure[]{\includegraphics[scale=0.9]{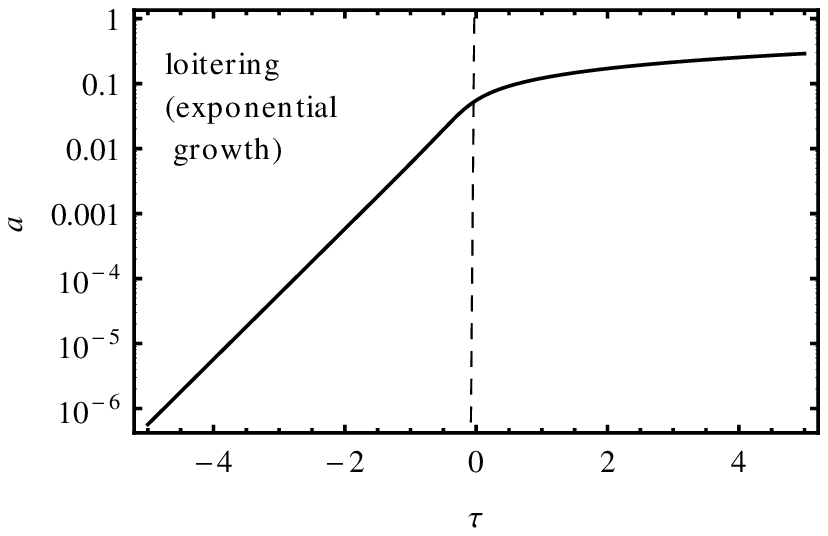}\label{subfig:a_kappapos_zoom2}}
\subfigure[]{\includegraphics[scale=0.9]{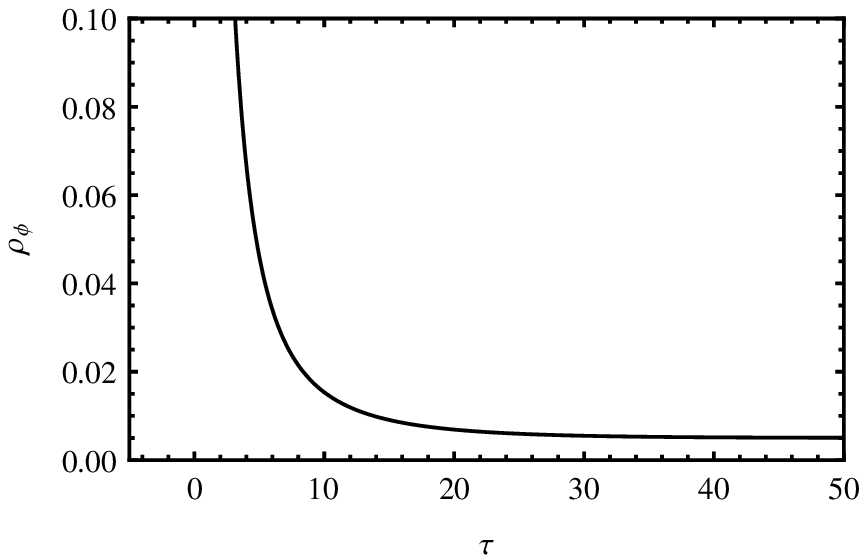}\label{subfig:rhophi_kappapos}}
\caption{The scale factor $a(\tau)$ is plotted as a function of 
cosmological time. For all the plots, we set $8\pi G=1$, $c=1$ and choose $\kappa =0.5, \, \alpha_T^2= 5.0, \, C_2=0.001,$ and $C_0=1.0025$. We choose an initial value $a=0.06$ at $\tau=0.06$. (a) The initial loitering state followed by a deceleration and late time acceleration are highlighted. (b) A zoomed version of (a) is shown to illustrate the loitering phase and the following transition to the deceleration phase. Note that the loitering phase also includes an acceleration, where the scale factor has an exponential growth ($a\sim a_0\exp(2\sqrt{2}c\tau/\sqrt{3\kappa})$). (c) The loitering phase is shown again by changing the linear scaling to a logarithmic scaling in the vertical axis. (d) The energy density ($\rho_{\phi}$) for the scalar field is plotted as a function of $\tau$.}
\end{figure}

Using Eqs.~(\ref{eq:a_X}) and (\ref{eq:tau_X}), we plot the scale factor $a$ as a function of the cosmological time ($\tau$) [Fig:\ref{subfig:a_kappapos}]. 
In the plot, we choose a value of $C_0$ such that we are able to
note an initial {\em loitering} phase (with an acceleration) 
followed by a decelerated expansion phase and then an accelerated expansion 
at late times. From Fig.~\ref{subfig:rhophi_kappapos}, we note that the 
effective energy density ($\rho_{\phi}$) decreases as $a$ increases and 
approaches a nonzero minimum value at large $a$. Though $\rho_{\phi}$ diverges 
at $a\rightarrow 0$, it takes an infinite time ($\tau \rightarrow -\infty$) 
to reach that point. Thus, the Universe is nonsingular. Such a nonsingular 
de Sitter inflationary phase is a common feature of the Universe 
with constant pressure, in $3+1$ EiBI cosmology. The accelerated expansion 
of the Universe at late times is due to the fact that 
at late times the effective pressure ($p_{\phi}$) is related to the 
energy density ($\rho_{\phi}$) as $p_{\phi}\approx -\rho_{\phi}c^2$. 
The decelerated expansion in between the two accelerated phases
is due to the fact that the repulsive nature of gravitating matter 
is less dominant
than its attractive character, during this phase.      

\begin{figure}[!htbp]
\centering
\subfigure[]{\includegraphics[scale=0.9]{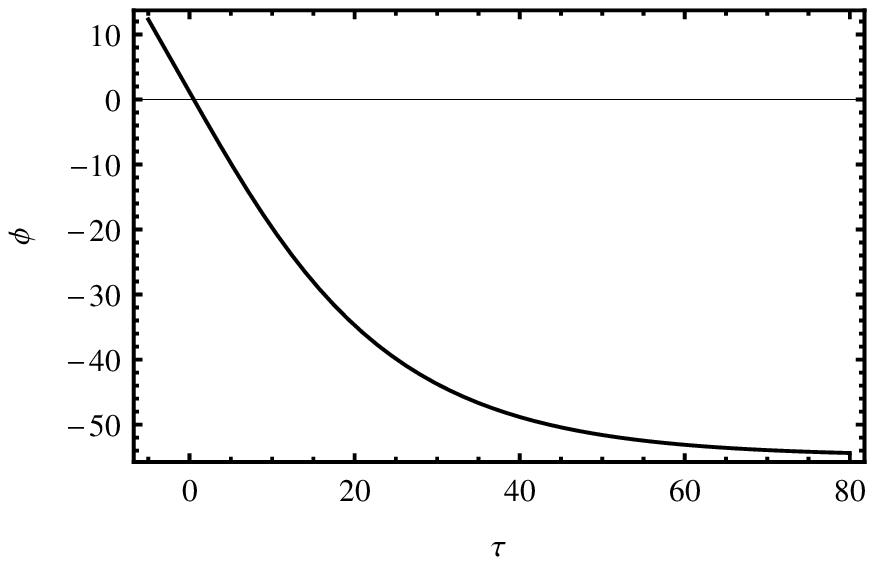}\label{subfig:phidecay_kappapos}}
\subfigure[]{\includegraphics[scale=0.9]{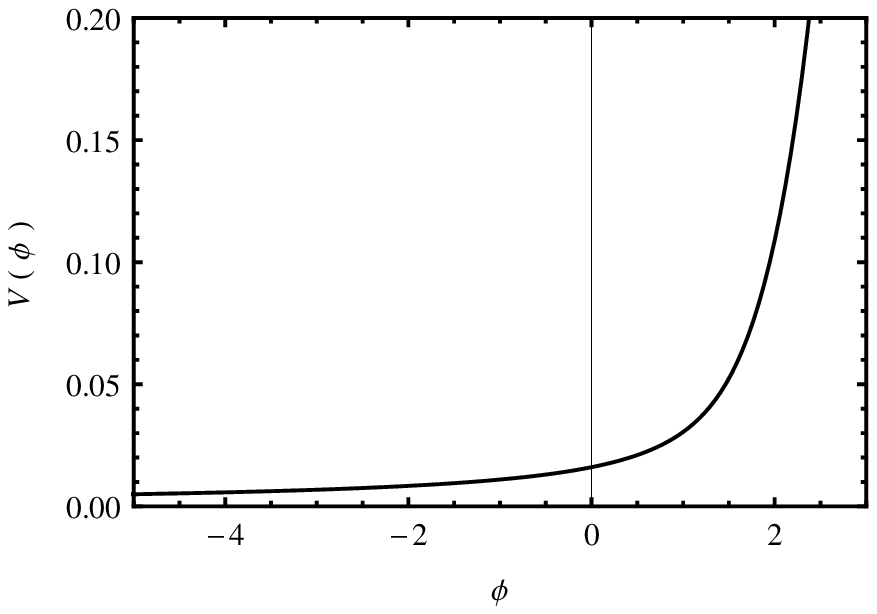}\label{subfig:pot_phidecay_kappapos}}
\subfigure[]{\includegraphics[scale=0.9]{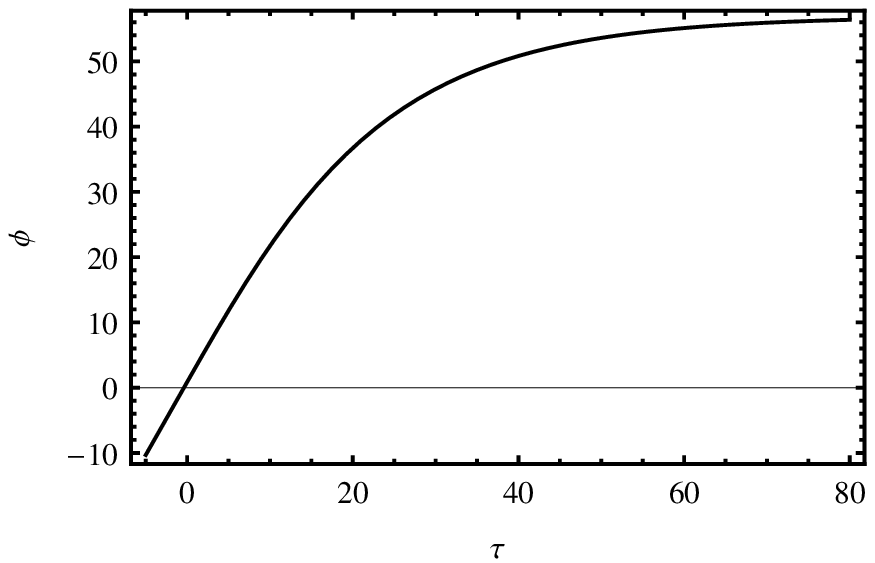}\label{subfig:phiincre_kappapos}}
\subfigure[]{\includegraphics[scale=0.9]{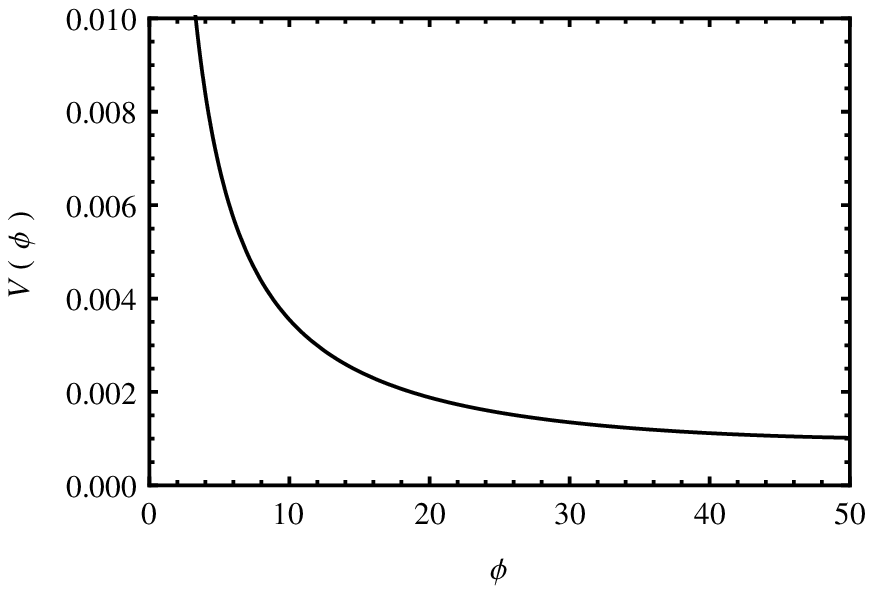}\label{subfig:pot_phiincre_kappapos}}
\caption{ (a) The scalar field ($\phi$) is plotted as a function of $\tau$. We assume $\dot{\phi}<0$ (decaying). (b) $\mathcal{V}(\phi)$ is plotted for the case $\dot{\phi}<0$. (c) $\phi(\tau)$ assuming $\dot{\phi}>0$ (growing). (d) $\mathcal{V}(\phi)$ for the $\dot{\phi}>0$ case. In all plots here, we choose an initial condition $\phi =1.0$ at $\tau=0.06$. }
\label{fig:phiandvphi}
\end{figure}

In Fig.~\ref{fig:phiandvphi}, we show the variation of the 
scalar field as a function of $\tau$, which results in such a cosmological 
solution. We also plot the associated potential $\mathcal{V}(\phi)$ as a 
function of $\phi$. An analytical expression for $\mathcal{V}$ as a function of $\phi$ is, unfortunately, not available due to the non-invertible nature of $\phi(a)$. We use Eqs.~(\ref{eq:phidot_assume}), (\ref{eq:pot_a}), 
and (\ref{eq:tau_X}) for the plotting. Here, the scalar field could be either decaying or 
growing in time for the same solution [see Eq.~(\ref{eq:phidot_assume})]. 
These  features are shown in the top and bottom panels respectively, 
with the associated potentials. However, in both situations, the potentials 
approach a nonzero minimum value and that occurs at late times of the Universe. 
   
During the loitering phase, the scale factor grows exponentially ($a\sim a_0\exp(2\sqrt{2}c\tau/\sqrt{3\kappa})$). On the other hand, it can be shown that during the period of inflation, the Universe expanded by a factor of $e^{60}$ in $10^{-32}$ seconds. Using this as input, the bound on $\kappa$ becomes $\kappa\lesssim 0.67\times 10^{-50}\, m^2$ which is greater than the Planck length square ($l_p=1.6\times 10^{-35}\, m$). However, the de Sitter expansion of the Universe at very late times is different from it and depends on the other BI parameter $\alpha_T^2$ ($a\propto \exp(\sqrt{8\pi G C_2 \alpha_T^2/3}\tau) $). Thus, in two different extreme regimes, the evolution of the scale factor depends on the two BI parameters independently. However, the intermediate phase depends on the product of these two parameters ($\kappa \alpha_T^2$).         

\subsection{$\kappa < 0$}
\begin{figure}
\centering
\includegraphics[width=0.7\textwidth]{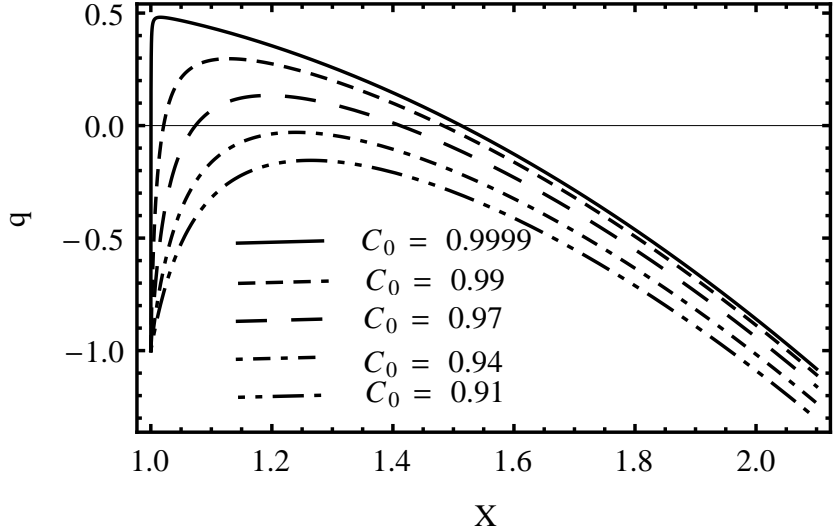}
\caption{The deceleration parameter ($q$) is plotted as a function of $X$ 
for $\kappa < 0$. In the plot, $X$ ranges from 1 to $\infty$ for $a\in (a_B ,\infty)$. $C_0$ [see Eq.~(\ref{eq:q_X})] takes different values for 
different curves in the plot. For $C_0 =0.9999, 0.99, 0.97$, $q$ has a 
transition from negative to positive to  negative values.  
For all values of $C_0$, $q\rightarrow -\infty$ for $X\rightarrow \infty$ 
or, $a\rightarrow a_B$. }
\label{fig:q_c0_kneg}
\end{figure}

For $\kappa < 0$, $a^3>0$ implies that $X\geq 1$ [see Eq.~(\ref{eq:a_X})]. For 
$X=1$, $a\rightarrow \infty$, but for $X\rightarrow \infty$ , $a^3\rightarrow {8\pi G |\kappa| C_2}/{C_0 c^2}$. Also, we note that $da/d\tau = 0$ at $X\rightarrow \infty$. So, for $\kappa <0$, there is a nonzero minimum scale factor $a_B= ({8\pi G |\kappa| C_2}/{C_0 c^2})^{1/3}$ at which  the Universe undergoes a {\em bounce}.  
In Fig.~\ref{fig:q_c0_kneg}, the deceleration parameter ($q$) is plotted as a function $X$ for different values of $C_0$. Since $\kappa <0$, so $C_0 <1$; we 
also assume $C_0 > 0$. In Fig.~\ref{fig:q_c0_kneg}, we note that $q\rightarrow -\infty$ as $a\rightarrow a_B$ (near the {\em bounce}). This is because at 
$a_B$, $da/d\tau=0$, but $d^2a/d\tau^2 $ has a nonzero finite value. 
However, in this case too, the Universe undergoes a de Sitter stage at 
late times ($q\rightarrow -1$ for $X\rightarrow 1$).

\begin{figure}[!htbp]
\centering
\subfigure[]{\includegraphics[scale=0.9]{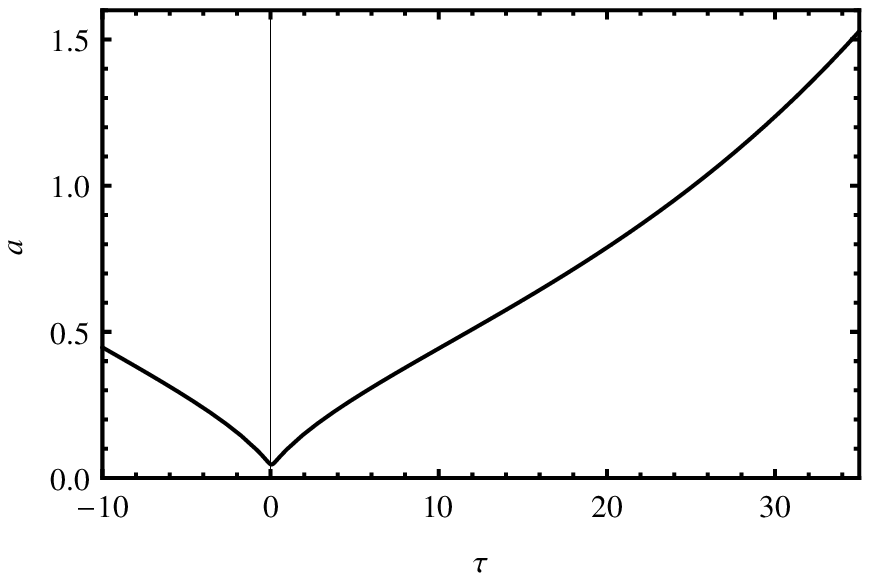}\label{subfig:a_kappaneg}}
\subfigure[]{\includegraphics[scale=0.9]{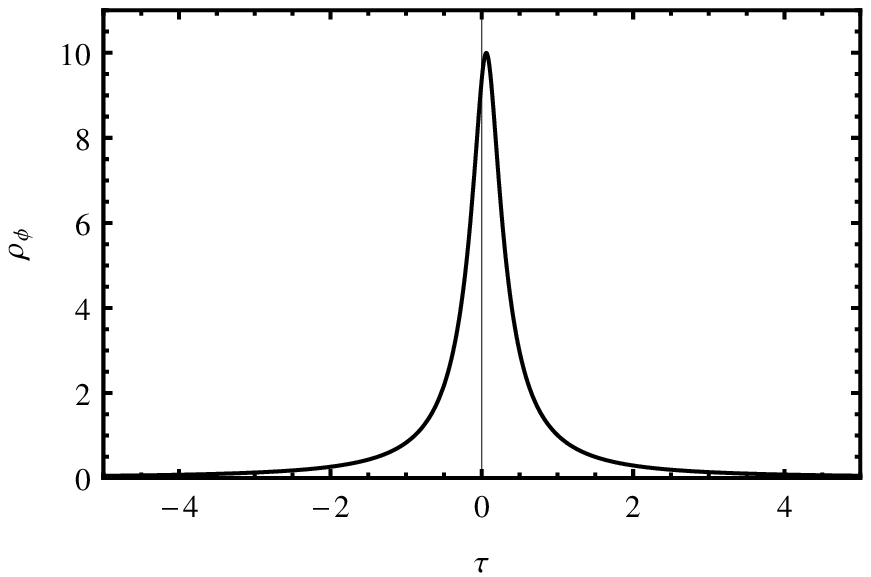}\label{subfig:rhophi_kappaneg}}
\caption{(a) The scale factor $a(\tau)$ is plotted as the function of the cosmological time ($\tau$) for $\kappa <0$. We set $8\pi G=1$, $c=1$ and choose $\kappa =-0.1, \alpha_T^2= 5.0, C_2=0.001,$ and $C_0=0.9995$. We choose an initial value $a=0.06$ at $\tau=0.06$. (b) 
The corresponding energy density ($\rho_{\phi}$) for the scalar field is also plotted 
as a function of $\tau$.}
\end{figure}

In Fig.~\ref{subfig:a_kappaneg}, we show the variation of the scale factor (
$a$) as a function of the cosmological time ($\tau$). We choose the parameter 
values in such a way that we see the {\em bounce} followed by decelerated 
expansion and finally the de Sitter stage at late times. The plot of the 
effective energy density ($\rho_{\phi}$) in the Fig.~\ref{subfig:rhophi_kappaneg} shows a regular profile throughout. Here we mention that the nonsingular 
{\em bouncing} early Universe is a common feature of EiBI cosmology. 
The BI scalar field provides the additional late time acceleration.

\begin{figure}[!htbp]
\centering
\subfigure[]{\includegraphics[scale=0.9]{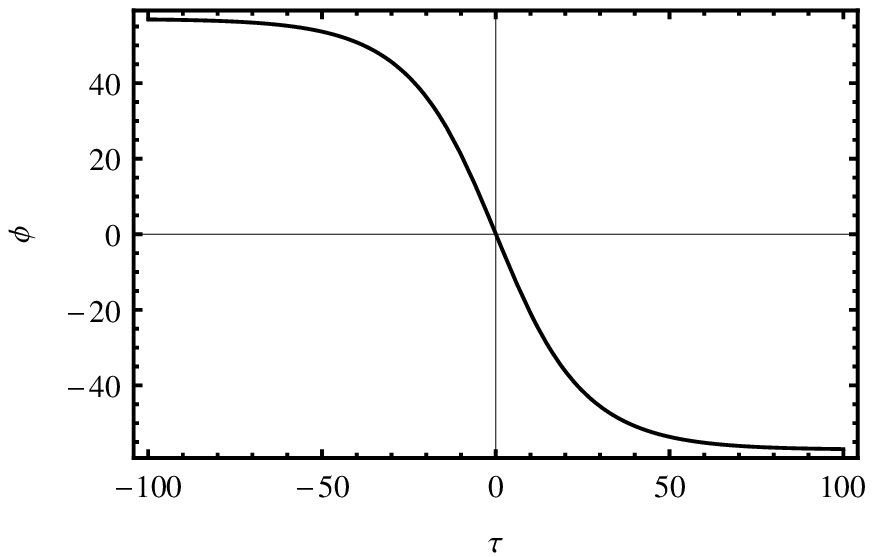}\label{subfig:phidecay_kappaneg}}
\subfigure[]{\includegraphics[scale=0.9]{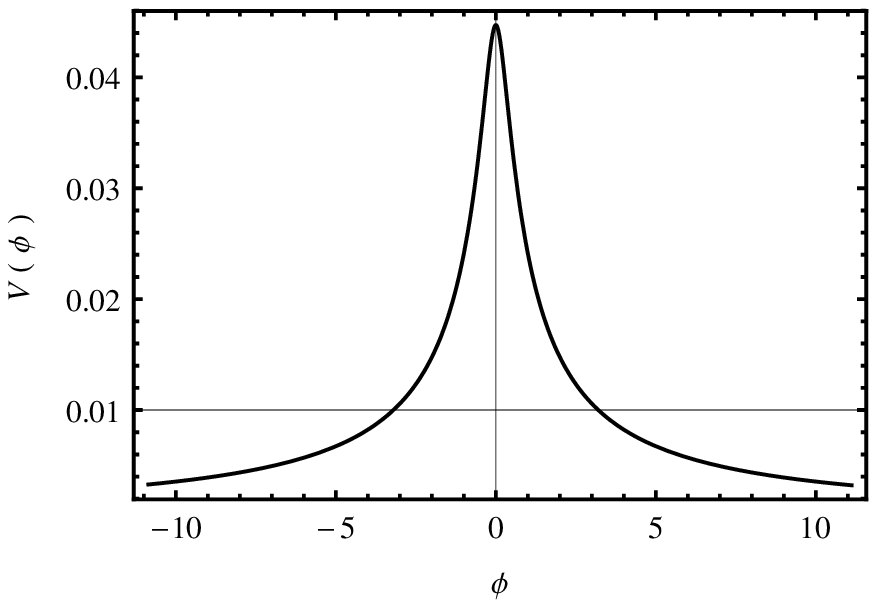}\label{subfig:pot_phidecay_kappaneg}}
\subfigure[]{\includegraphics[scale=0.9]{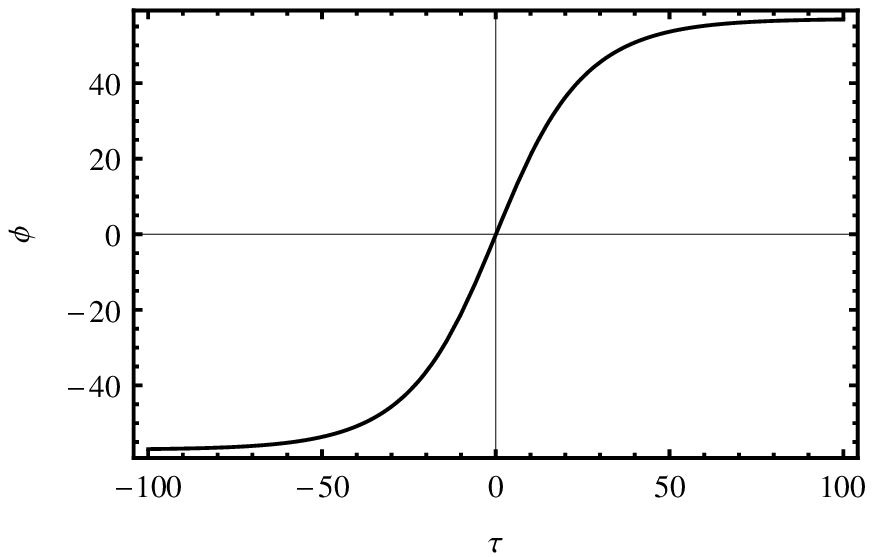}\label{subfig:phigrow_kappaneg}}
\subfigure[]{\includegraphics[scale=0.9]{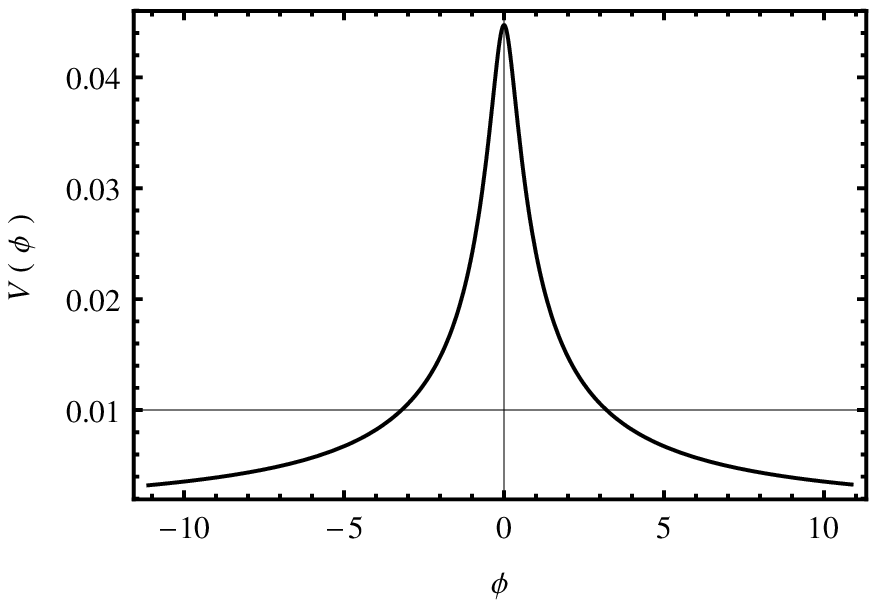}\label{subfig:pot_phigrow_kappaneg}}
\caption{(a) The scalar field ($\phi$) is plotted as a function of $\tau$. We assume $\dot{\phi}<0$ (decaying). (b) $\mathcal{V}(\phi)$ is plotted for the case $\dot{\phi}<0$. (c) $\phi(\tau)$ assuming $\dot{\phi}>0$. (d) $\mathcal{V}(\phi)$ for the case $\dot{\phi}>0$. In all plots here, we choose an initial condition $\phi =0.0$ at $\tau=0.06$. }
\label{fig:phiandvphi_kappaneg}
\end{figure}

In Fig.~\ref{fig:phiandvphi_kappaneg}, we plot the scalar field ($\phi (\tau)$) and the associated potential $\mathcal{V}(\phi)$. We note that, for both decaying and growing nature of the scalar field, the potential remains invariant.       

\section{Observational test of the solutions from fitting the supernova data}
So far we have found that the resulting cosmological solutions exhibit
an early-time accelerated expansion as well as a late-time acceleration. 
There can also exist a phase of decelerated expansion lying in between the
two acclerated phases. For $\kappa >0$, the early Universe grows 
exponentially (similar to an inflationary scenerio) and for $\kappa <0$, the early Universe undergoes a bounce.    
In this section, we fit the aforementioned solutions with the supernova 
data and test their viability in describing the evolution of the Universe 
at late times. However, we do not consider any other field 
as a part of the matter sector apart from the BI scalar field. Here, we follow 
the approach adopted in \cite{padmanabhan_tachyon1}, where the BI scalar field 
is treated as a single candidate having the capacity of exhibiting 
different equations of state at different scales and making a transition 
from $p=0$ (cold dark matter) at small scales to $p=-\rho c^2$ (dark energy) 
at large scales. The total energy density is, likewise, 
split into two parts: $(a)$ pressureless dust and $(b)$ dark energy, $i.e.$
\begin{eqnarray}
\rho_{\phi}&=&\rho_{DM} + \rho_{DE};\,\, p_{\phi}=p_{DM} + p_{DE}\nonumber \\
\mbox{where,}\,\,\, p_{DM}&=&0;\,\, p_{DE}=-\rho_{DE}c^2
\label{eq:dm_de_split}  
\end{eqnarray} 
Thus, from Eqs.~(\ref{eq:rho_phi}) and (\ref{eq:p_phi}), we get
\begin{equation}
\rho_{DM}=\frac{\mathcal{V}(\phi)\dot{\phi}^2}{c^4U\sqrt{1-\frac{\dot{\phi}^2}{c^2\alpha_T^2U}}}\hspace{0.1in};\hspace{0.2in} \rho_{DE}=\frac{\alpha_T^2\mathcal{V}(\phi)}{c^2}\sqrt{1-\frac{\dot{\phi}^2}{c^2U\alpha_T^2}}
\label{eq:rho_dm&de_phi}
\end{equation}
In our case, we get $\rho_{DM}=C_2/a^3$ and $\rho_{DE}= -p_{DE}/c^2=\alpha_T^2C_2$ (a constant) [see Eqs.~(\ref{eq:rho_a}) and (\ref{eq:p_a}) with $n=3$ and $C_1=\alpha_T^2$]. This is similar to $\Lambda$CDM cosmology. However, 
the cosmological constant $\Lambda = \alpha_T^2C_2$ is not an ad-hoc 
quantity but generated from the BI scalar field with a specific choice 
of $\dot{\phi}^2$ (or, alternatively with an equivalent choice 
of the potential function).

\subsection{$\kappa>0$}
Using Eqs.~(\ref{eq:a_X}) and (\ref{eq:tau_X}), we rewrite the Hubble 
function as
\begin{eqnarray}
H(z)&=&H_0\left[\frac{(\sqrt{X^2(z)+3}-X(z))(C_0-X(z))(X_0^2+3)}{(\sqrt{X_0^2+3}-X_0)(C_0-X_0)(X^2(z)+3)}\right]^{1/2}
\label{eq:H(z)_kappapos}
\end{eqnarray}  
where $z$ is the redshift defined as $a=1/(1+z)$. 
$H_0$ and $X_0$ are the present day values of the Hubble function 
and $X(z)$. The expressions for $X(z)$ and $H_0$ are given as
\begin{eqnarray}
X(z)&=& \frac{1-\tilde{a}_0^3(1+z)^3}{\sqrt{1+2\tilde{a}_0^3(1+z)^3}},\label{eq:X_a_z}\\
H_0&=&\frac{2c}{\sqrt{3\kappa}}\left[\frac{(C_0-X_0)(\sqrt{X_0^2+3}-X_0)}{X_0^2+3}\right]^{1/2}
\label{eq:H0}
\end{eqnarray}  
where, $\tilde{a}_0^3={4\pi G \kappa C_2}/{C_0 c^2}=(\sqrt{X_0^2+3}-2X_0)/(\sqrt{X_0^2+3}+X_0)$. Using the Eqs.~(\ref{eq:H(z)_kappapos}) and (\ref{eq:X_a_z}), we can define the luminosity distance for the observed supernova at the 
redshift $z$ as $D_L(z)=c(1+z)\int^z_0\frac{dz'}{H(z')}=cd_L(z)/H_0$, where 
$d_L(z)$ is the Hubble free luminosity distance.
Therefore, the Hubble free luminosity distance becomes
\begin{eqnarray}
d_L(z;X_0,C_0)&=&(1+z)\int^z_0 \frac{H_0dz'}{H(z')}\nonumber\\
&=& (1+z)\int^z_0\left[\frac{(\sqrt{X_0^2+3}-X_0)(C_0-X_0)(X^2(z')+3)}{(\sqrt{X^2(z')+3}-X(z'))(C_0-X(z'))(X_0^2+3)}\right]^{1/2}dz'
\label{eq:h_free_dist}
\end{eqnarray}
Using Eq.~(\ref{eq:h_free_dist}) as the model, we fit the supernova data. 
There are two parameters $C_0$ and $X_0$, which are both dimensionless. 
From the best fit parameter values (of $C_0$ and $X_0$), we can estimate the 
best fit values of $q_0$ (present day value of the deceleration parameter), 
$\Omega_{DM 0}$ (ratio of present day matter density to total energy density 
of the Universe), $\Omega_{DE 0}$ (for dark energy), $\kappa$, and $\alpha_T^2$.To carry out all of this, we write down the useful relations next. 
We use Eq.~(\ref{eq:q_X}) to evaluate $q_0$. We get the expression 
of $\Omega_{DM}(z)$ in terms of $X_0$ and $C_0$, [using Eqs.~(\ref{eq:rhophi_a}), (\ref{eq:a_X}), and (\ref{eq:H0})]
\begin{eqnarray}
\Omega_{DM}(z)&=& \frac{\rho_{DM}(z)}{\rho_{\phi}(z)}
= \frac{2C_0(\sqrt{X_0^2+3}-X_0)(\sqrt{X_0^2+3}-2X_0)(1+z)^3}{3\left[C_0\left(\sqrt{X^2(z)+3}-X(z)\right)^2-1\right]} 
\label{eq:omega_m0}
\end{eqnarray}
Further, $\Omega_{DE}(z)$ can also be expressed as 
\begin{eqnarray}
\Omega_{DE}(z)&=& \frac{\rho_{DE}}{\rho_{\phi}(z)}=1-\Omega_{DM}(z) 
\label{eq:omega_de0}
\end{eqnarray}
where $\Omega_{DM 0}=\Omega_{DM}(0)$ and $\Omega_{DE 0}=\Omega_{DE}(0)$.
We can evaluate $\alpha_T^2$ using
\begin{equation}
\alpha_T^2=\frac{\Omega_{DE 0}}{\Omega_{DM 0}}
\label{eq:alpha_T^2}
\end{equation}
Finally, the expression of $\kappa$ is
\begin{eqnarray}
\kappa &=& \frac{4c^2}{3H_0^2}\frac{(C_0-X_0)(\sqrt{X_0^2+3}-X_0)}{(X_0^2+3)} \label{eq:kappa_x0c0}
\end{eqnarray}
where, we use a prior value of $H_0=70\, km/sec/Mpc$. 

\noindent  To fit the Supernova data with our model, we follow the method 
used in \cite{supernova_fit}, wherein the authors have studied the expansion 
history of the universe upto a redshift $z=1.75$ using the 194 Type Ia 
supernovae (SNe Ia) data. However, in our case we use the more recent 
Union2.1 Compilation data \cite{suzuki}.
The observational dataset consists of the values of the 
distance modulus ($m_i(z_i)-M$) and redshifts $z_i$ with their 
corresponding errors. Each distance modulus is related to the 
corresponding luminosity distance $D_L$ of the SNe Ia by 
\begin{equation}
m(z)= M + 5log_{10}\left[\frac{D_L(z)}{Mpc} \right]+25
\label{eq:apparent_m}
\end{equation}

\noindent The observed distance modulus can be translated to $d^{obs}_L(z_i)$. For a given model $H(z;a_1,a_2,...,a_n)$, one can also theoretically predict the $d^{th}_L(z)$ using the Eq.~(\ref{eq:h_free_dist}). The best fit values of the 
model parameters ($a_1,a_2,....,a_n$) are estimated by minimizing 
the $\chi^2(a_1,a_2,....,a_n)$ which, in this case, is given 
by \cite{supernova_fit}
\begin{equation}
\chi^2(a_1,a_2,....,a_n)=\sum_{i=1}^{580}\frac{\left(log_{10}d_L^{obs}(z_i)-log_{10}d_L^{th}(z_i)\right)^2}{\left(\sigma_{log_{10}d_L(z_i)}\right)^2+\left(\frac{\partial log_{10}d_L(z_i)}{\partial z_i}\sigma_{z_i}\right)^2}
\end{equation} 
where $\sigma_z$ is $1\sigma$ redshift uncertainty of the data 
and $\sigma_{log_{10}d_L(z_i)}$ is the $1\sigma$ error of 
$log_{10}d_L^{obs}(z_i)$. The error in redshift $\sigma_z$ is estimated 
from the uncertainty due to peculiar velocities, $\Delta v=\Delta (cz)=500 \, km/s$, i.e. $\sigma_z=\Delta z= (500 \, km/s)/c$. 

\begin{figure}[!htbp]
\centering
\subfigure[]{\includegraphics[scale=1.3]{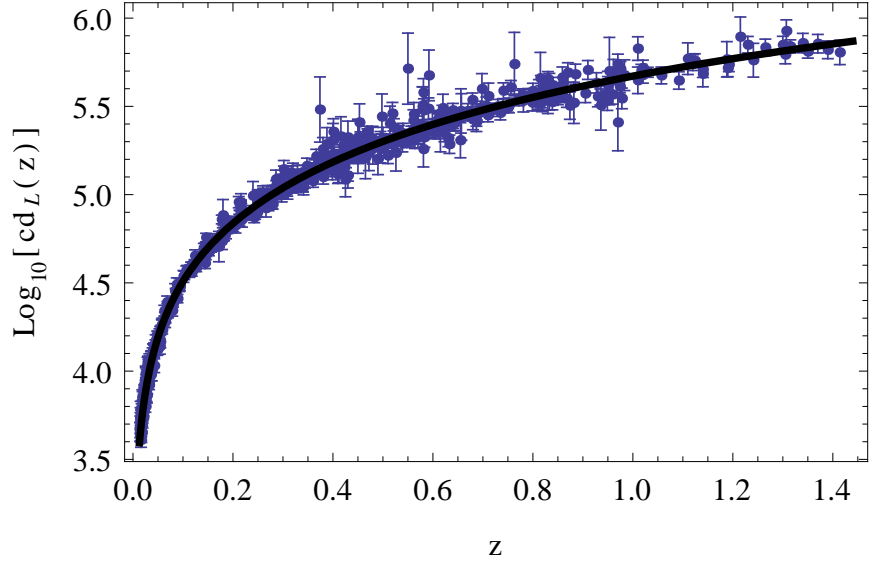}\label{subfig:ld_z_scalar}}\\
\subfigure[]{\includegraphics[scale=0.9]{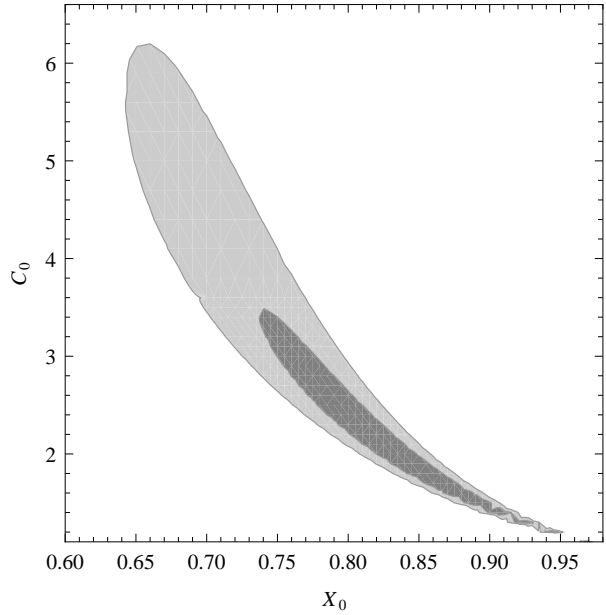}\label{subfig:errorplot_scalar}}
\caption{$(a)$ Observed SNe Ia (Union2.1 Compilation data) Hubble free luminosity distance ($d_L$) with the fitted curve (thick black solid line) is shown. 
In the plot we use $c=3\times 10^5\, km/s$. The best fit parameter values 
are $X_0=0.912$ and $C_0=1.372$; $\chi_{min}^2/d.o.f=0.861$. $(b)$ $1\sigma$ 
and $2\sigma$ error plots are shown in two-dimensional parameter space 
$(X_0-C_0)$.  }
\label{fig:supernova_fit}
\end{figure}

\noindent The resulting best fit parameter values are 

\begin{equation}
\boxed{X_0= 0.912 \hspace{0.2in}; \hspace{0.2in} C_0=1.372 \hspace{0.2in};
\hspace{0.2in} \chi^2_{min}= 497.926. }
\end{equation}
If $\chi^2_{min}/d.o.f=\chi^2_{min}/(N-n)\lesssim 1$ ($N$: number of data points, $n$: number of parameters), the fit is good and the data are consistent 
with the considered model $H(z; a_1,....,a_n)$. Here, $ \chi^2_{min}/d.o.f= 0.861$. In Fig.~\ref{subfig:ld_z_scalar}, the variation of the luminosity 
distance with respect to the redshift $z$ is shown for the best fit parameter 
values, along with the observed data points. In Fig.~\ref{subfig:errorplot_scalar}, we show the $1\sigma$ and $2\sigma$ confidence levels in the 
parameter space $(X_0-C_0)$. 

\begin{figure}
\centering
\includegraphics[width=0.7\textwidth]{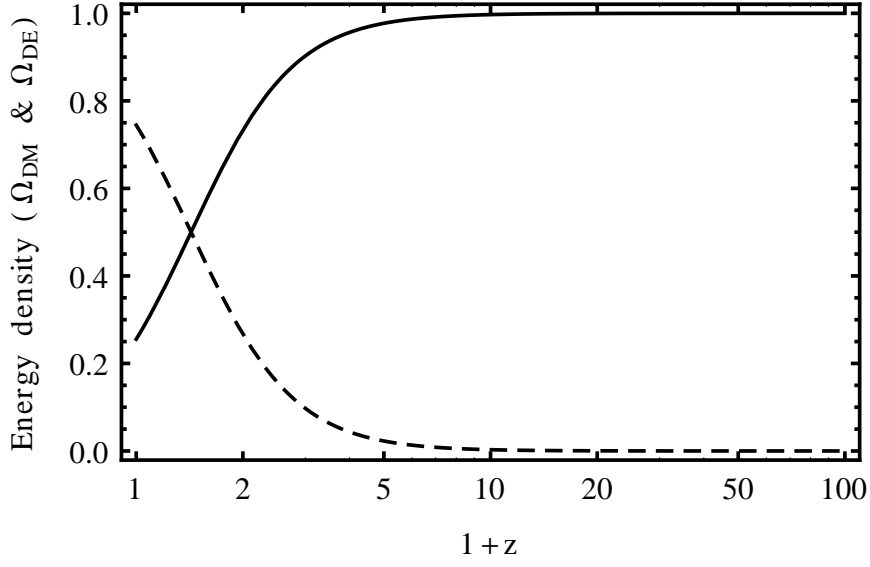}
\caption{The plot shows the evolution of $\Omega_{DM}$ (solid curve) and $\Omega_{DE}$ (dashed curve) with redshift ($z$) in our best-fit model for $\kappa>0$. The universe is completely dominated by matter beyond $z\sim 10$.}
\label{fig:energy_density_z}
\end{figure}

Using the best fit parameter values, the estimated values of $q_0$, $\Omega_{DM0}$, and $\Omega_{DE0}$  are 

\begin{equation}
\boxed{q_0=-0.605^{+0.026}_{-0.054}  \hspace{0.2in}; \hspace{0.2in} \Omega_{DM0}= 0.255^{+0.051}_{-0.021} \hspace{0.2in};
\hspace{0.2in} \Omega_{DE0}= 0.745^{+0.016}_{-0.051}}
\end{equation}
 These are in reasonably good agreement with $\Lambda$CDM cosmology \cite{suzuki}. Further, we plot $\Omega_{DM}(z)$ and $\Omega_{DE}(z)$ in Fig.~\ref{fig:energy_density_z}. This figure once again demonstrates the viability of
our model with observations, at least at late times.
 
\noindent We mention an important point here. 
Using the critical energy density of the Universe as obtained 
in the framework of GR (i.e. $\rho_{c0}=3H_0^2/{8\pi G}$), one can 
write down the present-day values of the density parameters corresponding to 
dark matter and dark energy as $\Omega'_{DM0}=8\pi G C_2/{3H_0^2}$ and $\Omega'_{DE0}=8\pi G C_2\alpha_T^2/{3H_0^2}$. 
Here, $\Omega'_{DM0}+\Omega'_{DE0}\neq 1$, as it should be
in a modified theory of gravity. 
In our case, with the best fit values of the parameters $C_0$ and $X_0$, we 
find that $\Omega'_{DM0}+\Omega'_{DE0}$ is very close (but not equal) to one 
($\Omega'_{DM0}=0.254$ and $\Omega'_{DE0}=0.741$). Therefore, irrespective of
our definition of the density parameters, our model for late-time evolution
seems to work well.       
 
The estimated value of the theory parameter for the EiBI gravity and the value of the parameter for the scalar field (matter) sector are 
\begin{equation}
\kappa = 3.1 \, (Gpc)^2 \hspace{0.2in}; \hspace{0.2in} \alpha_T^2 =2.9 .
\end{equation} 
 Using $\kappa$, $G$, and $c$, we can 
define mass ($[M]_{BI}$), time ($[T]_{BI}$), and length ($[L]_{BI}$). $[M]_{BI}=\sqrt{\kappa}G^{-1}c^2=7.13\times 10^{52}\, kg$, 
$[T]_{BI}=\sqrt{\kappa}/c=1.76\times 10^{17}\, s$, and $[L]_{BI}=\sqrt{\kappa}=1.76\, Gpc$.
\subsection{$\kappa<0$}
We also fit the {\em bouncing} ($i.e.\, \kappa<0$) solution with the 
same supernovae data. In this case, the expression for the luminosity distance 
function becomes
\begin{eqnarray}
d_L(z;X_0,C_0)&=& (1+z)\int^z_0\left[\frac{(\sqrt{X_0^2+3}-X_0)(X_0-C_0)(X^2(z')+3)}{(\sqrt{X^2(z')+3}-X(z'))(X(z')-C_0)(X_0^2+3)}\right]^{1/2}dz'
\label{eq:h_free_dist_kappaneg}\\
\mbox{where,}\quad~ X(z)&=& \frac{2+a_B^3(1+z)^3}{2\sqrt{1-a_B^3(1+z)^3}}\\
a_B^3 &=& \frac{4X_0-2\sqrt{X_0^2+3}}{\sqrt{X_0^2+3}+X_0}\label{eq:a_min}\\
\mbox{and,}\quad~ H_0&=&\frac{2c}{\sqrt{3\vert \kappa \vert}}\left[\frac{(X_0-C_0)(\sqrt{X_0^2+3}-X_0)}{X_0^2+3} \right]^{1/2} 
\end{eqnarray} 
Here, $X_0\geq1$ and $0<C_0<1$. The minimum scale factor ($a_B$) 
corresponds to the maximum redshift $z_{max}=1/a_B-1$. The best fit parameter 
values are $X_0=1.011$, $C_0=0.973$ with $\chi_{min}^2/d.o.f.=0.862$. Using 
the best fit value of $X_0$ in Eq.~(\ref{eq:a_min}), the maximum redshift 
becomes $z_{max}=3.51$. This is absurd as $z_{max}$ should be greater than 
the redshift corresponding to CMB radiation ($z\approx 1000$). 
Thus the  {\em bouncing} Universe model is ruled out though it fits well 
with the data. 

\section{Conclusions}
\noindent Let us now briefly summarize our findings.

\noindent We have looked at cosmologies in EiBI gravity 
with a Born-Infeld scalar (tachyon condensate) in the 
matter part of the total action. Thus, we have incorporated a 
Born-Infeld structure
in both the gravity and the matter sectors of the theory. 
We have  two control parameters; $\kappa$ of dimension $L^2$ for the 
gravity sector and the dimensionless $\alpha_T^2$ for the matter sector.

\noindent In our approach here, we have assumed a 
form of $\dot{\phi}^2$ instead of assuming a scalar potential. 
For a particular choice of a parameter $n$ (i.e. $n=3$), 
the problem reduces to a situation where the Universe is driven by a 
perfect fluid of constant negative pressure. We obtain the analytical 
solution for such a set up. For $\kappa>0$, the Universe undergoes a 
de Sitter expansion stage both at early and late times. In between, there  
could be a decelerated expansion depending on the value of 
$\kappa \alpha_T^2$. For $\kappa<0$, there is a difference in the picture 
through the occurrence of a {\em bounce} instead of the de Sitter expansion 
at early times. However, at late times, the Universe still undergoes 
the de Sitter expansion even for $\kappa<0$. 

\noindent Qualitatively, similar behaviour can also be achieved 
for $n>3$ in the form of $\dot{\phi}^2$ (Eq.~\ref{eq:phidot_assume}). 
For such a different choice, 
the effective pressure $p_{\phi}$ becomes bounded. This is named the 
Maximal Pressure State (MPS) in \cite{chokim}, which provides a de Sitter 
stage in the expansion history of the Universe at the early times 
for $\kappa>0$, in EiBI gravity. But for $\kappa<0$ and $n>3$, 
a similar {\em bounce} does occur 
supporting the  generic {\em bounce} character in EiBI cosmology. 
We also note that, in our case, the equation of a
state for the BI scalar (tachyon condensate) becomes 
$p_{\phi}\approx-\rho_{\phi}c^2$ at late times. Thus, at late times, 
the Universe undergoes a de Sitter expansion phase. 
In fact, it seems that we do not need any exotic matter source 
for producing late time 
acceleration in EiBI gravity, due to the above-mentioned property.

\noindent We have split the total energy-momentum tensor of the 
BI scalar (tachyon condensate) into two parts: one for 
dark matter ($p=0$) and the other as dark energy ($p=-\rho c^2$). In our 
special case ($i.e.$ $n=3$ in the equation of $\dot{\phi}^2$), 
the dark energy has constant negative pressure. Therefore, 
this is equivalent to $\Lambda$CDM cosmology though the effective 
cosmological constant is generated from the BI scalar. 
However, for $n>3$, we would have an evolving dark energy.

\noindent It may be noted that though we have a viable background cosmological model, issues such as inflation and reheating will have to be addressed in greater detail. We hope future investigations will throw light on these topics.
 
\noindent We have shown that the supernova data fit well with both the 
late time solutions for $\kappa>0$ and $\kappa<0$. However, 
we discard the $\kappa<0$ solution because our fit predicts an 
unacceptable value of the redshift where the Universe may undergo a bounce.
With a different choice of $\dot{\phi}$ and additional matter, 
it may be possible to introduce new parameters and obtain a viable $\kappa<0$ solution.  
Remarkably, the cosmological properties estimated from the supernova data 
fit of the $\kappa>0$ solution is as good as in $\Lambda$CDM cosmology. 
It is possible that instead of using the special case $n=3$, 
one may use the general form 
of $\dot{\phi}^2$ and keep $n$ as an additional fitting parameter, in order 
to figure out how dark energy evolves.       

\noindent Since we have analytical solutions which are not too complicated,
it may be worthwhile to attempt a study of cosmological perturbations
using this model as a background cosmology. Such a study with 
adequate observational tests and checks will surely 
help in establishing EiBI cosmology on a firmer footing, in future.



\bibliographystyle{apsrev4-1}
\bibliography{eibiscalar_cosmo_revised1}
\end{document}